\documentclass[10pt]{article}

\usepackage{amsmath, enumerate}
\usepackage{amssymb}
\usepackage{bm}
\usepackage{color}
\usepackage{graphicx}
\usepackage{multirow}
\usepackage[square, numbers, comma, sort&compress]{natbib}
\usepackage{verbatim}
\usepackage{hyperref}

\numberwithin{equation}{section}
\newtheorem{prop}{\indent{\sc Proposition}}
\newtheorem{rmk}{\indent{\sc Remark}}
\newtheorem{thm}{\indent{\sc Theorem}}
\newtheorem{defi}{\indent{\sc Definition}}

\newcommand{\benu}{\begin{enumerate}}
\newcommand{\eenu}{\end{enumerate}}

\newcommand{\beit}{\begin{itemize}}
\newcommand{\eeit}{\end{itemize}}
\newcommand{\be}{\begin{eqnarray}}
\newcommand{\bec}{\begin{center}}
\newcommand{\eec}{\end{center}}
\newcommand{\beo}{\begin{eqnarray*}}

\newcommand{\Bl}{\Bigl(}
\newcommand{\Br}{\Bigr)}

\newcommand{\ee}{\end{eqnarray}}
\newcommand{\eeo}{\end{eqnarray*}}
\newcommand{\mb}{\mathbb}
\newcommand{\mc}{\mathcal}

\newcommand{\n}{\nonumber}
\newcommand{\vp}{\varphi}

\newcommand{\la}{\lambda}

\newcommand{\no}{\nonumber}
\newcommand{\qed}{\hfill \rule{2mm}{2mm}}

\begin{document}

\title{Option Pricing in a Regime Switching Stochastic Volatility Model\thanks{This research was supported in part by the SERB MATRICS (grant MTR/2017/000543)}}
\author{Arunangshu Biswas\footnote{Novartis AG, Hyderabad India. Email: arunb12002@gmail.com}, Anindya Goswami\footnote{IISER, Pune 411008, India. Email: anindya@iiserpune.ac.in}~~\& Ludger Overbeck\footnote{Mathematics Institute, University of Giessen 35392, Germany. Email: Ludger.Overbeck@math.uni-giessen.de}}
\date{ }
\maketitle

\begin{abstract}
We consider a regime switching stochastic volatility model where the stock volatility dynamics is a semi-Markov modulated square root mean reverting process. Under this model assumption, we find the locally risk minimizing price of European type vanilla options. The price function is shown to satisfy a non-local degenerate parabolic PDE which can be viewed as a generalization of the Heston PDE. The related Cauchy problem involving the PDE is shown to be equivalent to an integral equation(IE). The existence and uniqueness of solution to the PDE is carried out by studying the IE and using the semigroup theory.
\end{abstract}


\section{Introduction}
The regime switching model of \cite{DKR} is an useful extension to the classical Black-Scholes model. There the market can take one of the finitely many hypothesized states/regimes for a random duration of time. The key market parameters, namely interest rate, mean growth rate, volatility remain constant at each regime but vary depending on regimes. Thus the asset price dynamics depends on an additional underlying process which is allowed to be a Markov pure jump process. However, the consideration of Markovian regime switching is not limited in generalizing Black-Scholes model only. The regime switching extension was successfully carried out for many other alternative models of asset price also. A comprehensive literature survey in this direction is beyond the scope of this paper. Nevertheless, we suggest the readers to see \cite{BAS, JR, SYL, SWH, R1, R2, R3, R5, R6, R7, R8, R10, R11} and references therein for the recent development in regime switching models.

In \cite{Goswami1, AJP}, the scope of the model in  \cite{DKR} is further extended to the case of semi-Markov regimes. Here the sojourn time at each regime is allowed to have a more general type of distribution than the class of exponential distributions. Such generalizations bring in some technicalities also. Since the state process is no more Markov, in order to obtain a Markov setting, one needs to augment the state process with the sojourn time process. The age or the sojourn time at a state grows like time deterministically in between two state transitions and jumps down to zero at the instant of next transition. Therefore the infinitesimal generator of the augmented process involves a non-local term and a first order differential operator only. For obvious reason, we would also see in this paper, such feature of generator w.r.t. the age variable, causes the option price equation to be a degenerate parabolic non-local partial differential equation (PDE). Thus the well-posedness of the related Cauchy problem is not straightforward. However, there is an advantage of using semi-Markov process in a regime switching model. Since the corresponding option price equation involves the age variable, one gets the scope of feeding in an empirical value of age variable in the equation to obtain an option price. On the other hand, in the Markov regime model, the option price is insensitive to the age value. Thus in view of increasing computational power and advancement of statistical inference techniques, the use of semi-Markov process for market regimes has become more legitimate choice.

The semi-Markov models and its variations have been suggested in the literature earlier also. For example, in \cite{Bulla} the authors use hidden semi-Markov models to model slow decay of the auto-correlation function in the squared daily returns of daily time series values. In \cite{Hunt} the financial market is a semi-Markov switching model where the holding time is distributed as a Gamma random variable. For other examples and applications one may see, for example \cite{Ladde} and references therein. It is known that a semi-Markov process may exhibit duration dependent transitions which a time homogeneous or a time in-homogeneous Markov chain do not. Empirical evidence of duration dependent transition of business cycles are also not uncommon, see, for example \citep{Chang}.

Since the market is incomplete for each of the models, mentioned above, the no arbitrage price is not unique. The local risk minimization is one among many other approaches adopted by the authors to price European type options in such incomplete markets. In this paper we address the theoretical option pricing problem via locally risk minimizing approach in a regime switching market model that can be viewed as a generalization of the Heston model \cite{heston}. The classical Heston model is a stock price model where the instantaneous volatility is itself a square root mean reverting process (also termed as Cox, Ingersoll and Ross (CIR) process). The long-run mean do appear in the stochastic differential equation(SDE) of the CIR process as a parameter. The SDE contains two other parameters as well, namely the speed of mean reversion, and the volatility of volatility. In a regime switching extension, all or some of the parameters and the drift coefficients of the asset prices are allowed to evolve as a finite state pure jump process. The rationale behind such modeling is to accommodate additional randomness parsimoniously with additional parameters for better fitting.

The regime switching generalization of Heston model was studied in \cite{SYL}. In \cite{SYL}, the switching was modeled by a finite state continuous time Markov chain with a given rate matrix. We refer to \cite{SWH,R10} for similar or further extensions of classical Heston model. The above literature survey is merely indicative but not exhaustive by any means. However, as per our knowledge, the consideration of semi-Markov switching of parameters in Heston model is absent in the literature. Needless to say, if semi-Markov switching is considered, the Markov subcase is not excluded from the consideration. Incorporating the duration dependent dynamics of regime switching is the main objective of the semi-Markov modulated continuous time models, as for example in \cite{Goswami1, AJP} which however fail to explain continuous variability of volatility parameter. In both \cite{Goswami1, AJP} the volatility is modeled as a pure jump process which is piece wise constant. The option pricing problem is not yet studied in market models which capture duration dependent continuous movement of stock volatility. The present model fills this gap and opens up possibility of further extensions. We have identified a set of sufficient conditions on the model parameters so that the proposed market model has no arbitrage. Furthermore, the existence of a minimal equivalent martingale measure is also shown.

For the sake of a broader scope, we have considered an abstract European type path independent option in which the contingent claim is a Lipschitz continuous function of terminal stock value. This class of claims include call, put, and butterfly, to name a few. We have shown using F\"ollmer Schweizer decomposition that the locally risk minimizing price of such a claim can be expressed as a solution to an appropriately chosen Cauchy problem. Here the corresponding PDE can also be viewed as a generalization of the Heston PDE. This Cauchy problem does not possess a closed form solution. Moreover, existence of a classical solution does also not follow immediately from the existing results in the literature. In this paper, we establish the existence and uniqueness of the classical solution to the problem. In this part, we have used treatment of semigroup theory for solving the abstract Cauchy problems. At first the price equation is shown to have a continuous mild solution satisfying an integral equation. Then by studying the integral equation, it is proved that the mild solution is sufficiently smooth and solves the PDE classically.

This paper is arranged as follows. In Section 2 the description of underlying asset price dynamics is presented. Next we propose the option price equation in Section 3. In this section, we show the existence and uniqueness of the classical solution to the equation. In Section 4 we show that the solution is indeed the locally risk minimizing price of the relevant European option. We conclude this paper with some remarks in Section 5.

\section{Model Description}
This section is divided in four subsections. The dynamics of the price of risky asset is presented in the first subsection. The second one provides detailed formulation of the semi-Markov process. No-arbitrage of the model is established in the third part whereas the fourth subsection proves existence of a minimal martingale measure.
\subsection{Asset Price Dynamics}
If $S=\{S_t\}_{t\ge 0}$ denotes the price dynamics of the risky asset, then according to the classical Heston model $S$ solves the following SDE
\be \left.
\begin{array}{lll}
 dS_t &=& \mu S_t dt + \sqrt{V_t} S_t d W_t^1, \ S_0 >0\\
 d V_t &=& \kappa(\theta - V_t ) dt + \sigma \sqrt{V_t} dW_t^2, \ V_0 >0. \label{Heston orig}
 \end{array}
 \right\}
 \ee
and $d\langle W^1,W^2 \rangle_t = \rho dt$ for some $\rho \in [-1,1]$ where $\theta(>0)$ is termed as the long run mean of volatility $V=\{V_t\}_{t\ge 0}$. Moreover the parameters $\kappa$ and $\sigma$ are positive constants and the Feller condition $\sigma^2< 2\kappa \theta$ holds true. While the Feller condition assures positivity of $V$, another condition, namely $ \sigma \le \frac{\kappa}{(2\rho + \sqrt 2)^+}$ assures square integrability of $S$, (see  \cite{AP}). In this paper we allow the drift coefficient $\mu$, the speed parameter $\kappa$, the volatility of volatility $\sigma$ and the parameters $\theta$ to evolve depending on an underlying semi-Markov process $X=\{X_t\}_{t\ge 0}$ with finite state space $\mc{X}= \{1,2 \ldots, k\} \subseteq \mb{R}$  and instantaneous transition rate functions $\lambda=\{ \lambda_{ij}:[0,\infty)\to (0,\infty) \mid i \neq j \in \mc{X}\}$. In addition to the risky asset, i.e. stock, we assume that the market includes another locally risk free asset with price $B_t$ per unit of money at time $t$. At time $t$ the instantaneous interest rate $r_t$ of the risk-free asset is $r(X_t)$. To be more precise, the model of stock and the money market instrument price is given below
\be
&&\left.
\begin{array}{lll}
 dS_t &=& \mu_t S_t dt + \sqrt{V_t} S_t d W_t^1, \ S_0 >0\\
 d V_t &=& \kappa_t(\theta_t - V_t) dt + \sigma_t \sqrt{V_t} dW_t^2, \ V_0 >0. \label{Heston}
\end{array}
\right\}\\
\n && \  dB_t \quad = \quad r_t B_t dt
\ee
where $\mu_t = \mu(X_t)$,  $\theta_t = \theta(X_t)$, $\kappa_t = \kappa(X_t)$, and $\sigma_t = \sigma(X_t)$. We assume that $\kappa(i), \sigma(i)$, $\theta(i)$, $\mu(i)$, and $r(i)$ are positive constants with $\mu(i) \ge r(i)$ such that
\begin{itemize}
\item[{\bf(A1)}] $\displaystyle \sigma(i) < \min\left(\sqrt{2 \kappa(i) \theta(i)+\rho^2(\mu(i)-r(i))^2}-\rho(\mu(i)-r(i)), \ \frac{\kappa(i)}{(2\rho + \sqrt 2)^+}\right)\ \forall \ i \in \mc{X}.$
\end{itemize}
The processes $W^1$ and $W^2$ are as before and taken to be adapted to the filtration $\{\mc{F}_t\}_{t\ge 0}$ satisfying the usual hypothesis and are defined on the probability space $(\Omega, \mc{F}, P)$. The semi-Markov process is also taken to be $\{\mc{F}_t\}_{t\ge 0}$ adapted and independent to both $W^1$ and $W^2$. We note that equation (\ref{Heston}) differs from the classical Heston model (\ref{Heston orig}) only by the parameters $\mu, \kappa, \sigma$ and $\theta$ which are non-explosive, finite-state pure jump processes. Therefore the positivity of $V$ and square integrability of $S$ follow similarly under (A1). Note that (A1) is stringent than required. Indeed a weaker assumption
$\sigma(i) < \min\left( \sqrt{2 \kappa(i) \theta(i)}, \frac{\kappa(i)} {(2\rho + \sqrt 2)^+}\right)$ for all $i$ is sufficient for positivity of $V$ and square integrability of $S$. However, (A1) is assumed so as to ensure positivity of another CIR process to be considered in Section 3 which would be clarified in due course. Henceforth we assume (A1) throughout.

\subsection{Formulation of Semi-Markov Process}
Let $T_n$ be the time instant of $n$th transition of the pure jump process $X$ and $T_0=0$. The age value at time $t$ is given by $t-T_{n(t)}$ where $n(t)$ denotes the number of transitions till time $t$. Also let $\tau_{n+1}:= T_{n+1}-T_n$ denote the time duration between $n^{th}$ and $(n+1)^{st}$ transitions. The instantaneous transition rate function $\lambda_{ij}: [0, \infty) \to [0, \infty), \forall i \neq j \in \mc{X}$ of the semi-Markov process $X$, if exists, is given by
$$\lambda_{ij}(y):=\lim_{\delta \to 0} \frac{1}{\delta} P\left[ X_{T_{n+1}}=j, \tau_{n+1}\in (y, y+\delta) |X_{T_n}=i, \tau_{n+1} >y \right]$$
where the conditional probability on the right side is independent of choice of $n$. We consider the class of semi-Markov processes which admits the above limit for each nonnegative $y$. In addition to this we further assume that
\begin{itemize}
\item[{\bf(A2)}](i) For each $i\neq j \in \mc{X}$, $\lambda_{ij}:[0,\infty)\to (0,\infty)$ is a continuously differentiable function. \\
(ii) If $\la_i(y):=\sum_{j\in \mc{X}\setminus \{i\}}\lambda_{ij}(y)$ and $\Lambda_i(y):=\int_0^y\la_i(y) dy$, then $\lim_{y\to \infty}\Lambda_i(y) =\infty$.
\end{itemize}
\begin{rmk}
\begin{enumerate}[i.]
  \item Note that for the special case of a finite-state continuous time Markov chain the instantaneous transition rates turn out to be positive constants. Therefore, assumptions in (A2) include all finite-state continuous time Markov chains.
  \item We demanded existence of transition rate on $[0,\infty)$. This is more stringent than what we require. Since our only concern is to price a European option with maturity $T$, say, we require existence of $\lambda$ on $[0,T]$ only.
  \item By defining $\tau_1 =T_1$, we have tacitly assumed that $X$ has no memory at time $t=0$. This can also be relaxed by setting $T_0$ a negative value.
\end{enumerate}
\end{rmk}
Here we recall the following results from \cite{AJP}.
\begin{prop}\label{prop} Given a collection $\lambda=\{ \lambda_{ij}:[0,\infty)\to (0,\infty) \mid i \neq j \in \mc{X}\}$ of bounded measurable maps, satisfying (A2(ii)), the following hold.
\begin{enumerate}[i.]
	\item Given an $\mc{F}_0$ measurable $X_0$, and a non-positive constant $T_0$, there exist a finite interval $\mb I$ and piecewise linear maps $h_{\lambda}$ and $g_{\lambda}$ on $\mc{X}\times [0,\infty)\times \mb I$ such that the system of coupled stochastic integral equations
\be \left.
\begin{array}{lll}
	X_t&=\displaystyle{ X_0+\int_{0}^{t}\int_{\mb{I}}h_{\lambda}(X_{u-},Y_{u-},z)\,\wp(du,dz)},\\
	Y_t&=\displaystyle{ (t-T_0) -\int_{0}^{t}\int_{\mb{I}}g_{\lambda}(X_{u-},Y_{u-},z)\,\wp(du,dz)},\label{ydef}
	\end{array}
\right\}
\label{Y}
\ee
where $\wp$, a Poisson random measure with uniform intensity, is independent of $X_0$ and adapted to $\{\mc{F}_t\}_{t\ge 0}$, has a strong rcll solution $(X,Y)$ such that $X$ is a semi-Markov process on $\mathcal{X}$ with $\lambda$ as instantaneous transition rate functions and $Y$ is the age process.
\item The infinitesimal generator $\mathcal{A}$ of $(X,Y)$ is given by $\mathcal{A}\vp(i,y) = \frac{\partial\vp}{\partial y}(i,y) + \sum_{j \neq i} \lambda_{ij}(y) \Bigl( \vp(j,0) - \vp(i,y)\Bigr)$ for every continuously differentiable function $\vp: \mathcal{X}\times [0,\infty)\to \mb{R}$.
\item Consider $F:[0,\infty)\times \mc{X}\rightarrow [0,1]$, defined as $F(y|i):=1-e^{-\Lambda_i(y)}$, where $\Lambda_i$ is as in (A2)(ii). Then under (A2)(i) $F(\cdot \mid i)$ is a twice continuously differentiable function and is the conditional c.d.f of the holding time of $X$ given that the present state is $i$.
\item Let for each $j\neq i$, $p_{ij}(y):=\frac{\la_{ij}(y)}{\la_i(y)}$ with $p_{ii}(y)=0$ for all $i$ and $y$.  Then $p_{ij}(y)$ denotes the conditional probability of transition to $j$ given that the process transits from $i$ at age $y$.
\item Let $f(y|i):=\frac{d}{dy}F(y|i)$, then $\la_{ij}(y) = p_{ij}(y)\frac{f(y|i)}{1-F(y|i)}$ hold for all $i\neq j$.
\end{enumerate}
\end{prop}
We also assume the following irreducibility condition.
\begin{itemize}
\item[{\bf(A2)}](iii) Set $\hat{p}_{ij}:=\int_0^\infty p_{ij}(y) dF(y|i)$. The matrix $(\hat{p}_{ij})_{k\times k}$ is irreducible.
\end{itemize}
Note that $(\hat{p}_{ij})_{k\times k}$ denotes the transition probability matrix of the embedded discrete time Markov chain.

\subsection{No Arbitrage of Market Model}
Henceforth we consider the SDE (\ref{Heston}) along with (\ref{Y}). Clearly $S, V, X, Y$ satisfying (\ref{Heston})-(\ref{Y}) are adapted to $\{ \mc{F}_t \}_{t \ge 0}$.
\begin{defi}\label{def1}
An admissible strategy is defined as a predictable process $\pi=\{\pi_t=(\xi_t,\varepsilon_t)\}_{0\leq t\leq T}$ if it satisfies the following conditions
	\begin{itemize}
		\item[(i)] $\xi:=\{\xi_t\}_{0\leq t\leq T}$ is square integrable w.r.t. $S$, that is,
 $$ E \Bigl(\int_0^T {\xi_t}^2 d \langle S\rangle_t \Bigr) <\infty;$$
		\item[(ii)] $E(\varepsilon^2_t)<\infty~\forall~ t\in [0,T]$; and
		\item[(iii)] if $J_t(\pi):=\xi_t S_t +\varepsilon_t B_t$, then ${P}(J_t(\pi)\geq -a,~\forall~ t\in [0,T])=1$ for some real $a$.
	\end{itemize}
	\end{defi}
It is evident that $J(\pi):= \{J_t(\pi)\}_{0\le t\le T}$ as in Definition \ref{def1} (iii) denotes the portfolio value process corresponding to the portfolio strategy $\pi$.

\begin{rmk}
The existence of an equivalent martingale measure(EMM) is a sufficient condition (see Theorem VII.2c.2 of \cite{SH}) for no arbitrage(NA) under admissible strategies. However, in general a stochastic volatility model need not admit an EMM. See \cite{SIN} for a detailed discussion on stochastic volatility models lacking EMM. We establish existence of an EMM under following assumption which we use throughout this paper.
\end{rmk}
\begin{itemize}
\item [{\bf(A3)}] Let $c := \frac{1}{2} \max_{i \in \mc{X}}(\mu(i) - r(i))^2$. Then $E(e^{ \int_{0}^{T} \frac{1}{V_s} ds})^c < \infty$.
\end{itemize}

\begin{thm}\label{NA}
Under (A1) and (A3) the above market model has no arbitrage under admissible strategies(as in Definition \ref{def1}).
\end{thm}
\textbf{Proof:} Consider $Z_t := \exp\left( - \int_{0}^{t} \frac{\mu_u - r_u}{\sqrt{V_u}} dW_u^1 - \frac{1}{2} \int_{0}^{t} \frac{(\mu_u - r_u)^2}{V_u} du\right)$. Using (A3) and the Novikov's condition (see, for example \cite{Pascucci}) we have that $Z=\{Z_t\}_{t\ge 0}$ is a $P$-martingale. This implies that $E_{{P}}(Z_T) = E_{{P}}(Z_0) = 1$. Now define a measure $\tilde{P}$ such that
\begin{equation}\label{PZ}
d \tilde{P} = Z_T d {P}.
\end{equation}
Therefore $\tilde{P}$ is a probability measure equivalent to ${P}$. From the Girsanov-Meyer Theorem (see, for example, Theorem III.39 \citep{PRO}) $\tilde{W}=\{\tilde{W}_t\}_{t\ge 0}$ is a $\tilde{P}$-martingale where $\tilde{W}_t := {W}^1_t + \frac{\mu_t - r_t}{\sqrt{V_t}}$. Therefore
\beo
d S_t &=&  S_t \Bigl( \mu_t dt  + \sqrt{V_t} d W_t^1\Bigr) \\
&=& S_t \Bigl( \mu_t dt + \sqrt{V_t} (d \tilde{W}_t - \frac{\mu_t - r_t}{\sqrt{V_t}} dt )  \Bigr) \\
&=& S_t \Bigl( r_t dt + \sqrt{V_t} d\tilde{W_t} \Bigr).
\eeo
Again since $r_t$ is the drift of $\{B_t\}_{t\ge 0}$ process, from above equation we conclude that the discounted stock price process $S^*=\{S^*_t\}_{t\ge 0}$, given by $S_t^* := \frac{S_t}{B_t}$, is a local martingale under $\tilde{P}$. Furthermore, from (A1), $S^*$ is square integrable and therefore a $\tilde{P}$-martingale. Or in other words, $\tilde{P}$ is an equivalent martingale measure. Hence the assertion of the theorem follows using Theorem VII.2c.2 of \cite{SH}. \qed

\subsection{Minimal Martingale Measure}
Note that since the stock price is dependent on an additional semi-Markov process which is non-tradable, the market model is incomplete. This implies that there is no unique risk neutral measure and therefore multiple no-arbitrage price for a terminal payoff.

Here we show that $\tilde{P}$, as constructed in (\ref{PZ}), is the Minimal Martingale Measure (MMM). Or in other words, we wish to show that every martingale $\hat{M}$ under $P$ which is orthogonal to $\int_0^\cdot \sqrt{V_s}S_s dW^1_s$, the martingale part of $S$ (as in \ref{Heston}) is also a martingale under $\tilde{P}$. Now since $V$ and $S$ are positive the above orthogonality implies that $\hat{M}$ is orthogonal to $W^1$. However the density process $Z$, as in the proof of Theorem \ref{NA}, solves the equation
$$Z_t=1- \int_0^t Z_u \frac{\mu_u-r_u}{ \sqrt{V_u}}dW^1_u.$$
Thus being an integral w.r.t. $W^1$, the process $Z$ has zero quadratic covariation with $\hat{M}$, i.e., $\langle\hat{M},Z\rangle =0$.

On the other hand the general Girsanov-Meyer Theorem asserts that if $Z$ is the density process, then
\be
\tilde{M} :=\hat{M} - \int_0^\cdot Z_u^{-1}d\langle\hat{M},Z \rangle_u. \label{Girsanov}
\ee
is a martingale under $\tilde{P}$. As $\langle\hat{M},Z\rangle =0$, from (\ref{Girsanov}), $\tilde{M} =\hat{M}$. Therefore, $\hat{M}$ is itself a martingale under $\tilde{P}$, as claimed.
Thus following the notion of locally risk minimizing pricing (see \cite{S2}), we conclude the following.
\begin{thm} Let $H:\Omega \to \mb{R}$ be a $\mathcal{F}_T$ measurable square integrable function. The locally risk minimizing fair price $C_t$ at time $t\le T$ of a terminal pay-off $H$ with terminal time $T$ is given by
$$C_t = \tilde{E}[e^{-\int_t^T r(X_u)du} H | \mathcal{F}_t]$$
where $\tilde{E}$ is the expectation w.r.t. the MMM  $\tilde{P}$ as above.
\end{thm}
The goal of this paper is to find an expression of $C_t$ as a deteministic function of $S_t, V_t, X_t, Y_t$ when $H$ is a given function of $S_T$. We accomplish that, using the F\"ollmer-Schweizer decomposition approach under certain conditions on the model parameters and the pay-off.
\section{Price equation}
Consider $\mc{D} = \{ (t,s,v, i,y) \in (0,T) \times (0,\infty) \times (0, \infty) \times \mc{X} \times (0,T)| y < t\}$. From (\ref{Heston}) and using Proposition \ref{prop} (ii) we deduce the infinitesimal generator $\mc{L}$ of the Markov process $(S,V,X,Y)=\{(S_t, V_t, X_t, Y_t)\}_{t\ge 0}$ as
\begin{eqnarray*}
  \mc{L}\vp (s,v,i,y)&=& \left(\mu(i) s \frac{\partial}{\partial s} + \kappa(i)( \theta(i) - v )\frac{\partial}{\partial v} + \frac{1}{2} v s^2 \frac{\partial^2}{\partial s^2} +  \frac{1}{2} \sigma^2(i) v \frac{\partial^2 }{\partial v^2} + \rho \sigma(i) v s \frac{\partial ^2 }{\partial s \partial v}  \right)\vp(s,v,i,y)\\
   &&  + \frac{\partial}{\partial y}\vp(s,v,i,y)  +\sum_{j \neq i} \lambda_{ij}(y) \Bigl( \vp(s,v,j,0) - \vp(   s,v,i,y) \Bigr)
\end{eqnarray*}
where $\vp$ is any smooth function on $\mc{D}$ with compact support.
We consider a Lipschitz continuous pay-off function $K:[0,\infty)\to [0,\infty)$ such that $|K(s)- c_1s |\le c_2$ for some nonnegative constants $c_1, c_2$. The pay-off functions associated to the call, put and butterfly options are included in the above class. We state a Cauchy problem:
\be\label{2a}
\left(\frac{\partial}{\partial t} +\mc{L}+ (r(i)-\mu(i))s\frac{\partial }{\partial s} - \rho \sigma (i) (\mu(i)-r(i))\frac{\partial }{\partial v}\right)\vp (t,s,v,i,y) &=& r(i) \vp (t,s,v,i,y)
\label{main eqn.}
\ee
in $\mc{D}$ with the terminal condition
\be \label{2b}
\vp(T,s,v,i,y) = K(s).
\ee
It is important to note that the second order partial derivative w.r.t. $y$ variable is absent. In particular (\ref{2a}) is a linear, non-local, degenerate parabolic system of PDEs. The non-locality is due to the presence of the term $\vp(t,s,v,j,0)$. Furthermore, the terminal data is merely Lipschitz continuous, thus need not be differentiable w.r.t. the space variables. Hence the existence of classical solution is not known a priori.

In this section, we establish the existence and uniqueness of the classical solution to (\ref{2a})-(\ref{2b}) in the class of functions with at most linear growth. To this end we consider the strong solution to another set of SDEs
\begin{eqnarray}
\left.
\begin{array}{lll}
 d\hat{S}_t &=& r(X_t) \hat{S}_t dt + \sqrt{\hat{V}_t} \hat{S}_t d W_t^1, \\
 d \hat{V}_t &=& \kappa(X_t)(\hat{\theta}_t - \hat{V}_t) dt + \sigma(X_t) \sqrt{\hat{V}_t} dW_t^2,
\end{array}
\right\}
\label{S.2}
\end{eqnarray}
where $\hat{\theta}_t:= \theta(X_t)- \frac{\rho \sigma(X_t)(\mu(X_t)-r(X_t))}{\kappa(X_t)}$ and $(X,Y)$, $W^1$, $W^2$ are as before. It is easy to see that under (A1),
$$\sigma(i) < \sqrt{2 \kappa(i) \left( \theta(i)- \frac{\rho \sigma(i)(\mu(i)-r(i))}{\kappa(i)}\right)}
$$
for all $i$. Hence $\hat{V}$ is positive almost surely and since $\sigma(i) <  \frac{\kappa(i)}{(2\rho + \sqrt 2)^+}$, $\hat{S}$ is square integrable. By comparison with (\ref{Heston}) the infinitesimal generator $ \hat{\mc{L}}$ of the Markov process $\{(\hat{S}_t, \hat{V}_t, X_t, Y_t)\}_{t\ge 0}$ is given by
$$ \hat{\mc{L}} = \mc{L}+ (r(i)-\mu(i))s\frac{\partial }{\partial s} - \rho \sigma(i)(\mu(i)-r(i))\frac{\partial }{\partial v}
$$
where $\mc{L}$ is the infinitesimal generator  of the Markov process $\{(S_t, V_t, X_t, Y_t)\}_{t\ge 0}$. Hence in view of Proposition \ref{prop}(ii), (\ref{2a}) can be rewritten as
\begin{equation}\label{3c}
\left(\frac{\partial }{\partial t} + \hat{\mc{L}}\right)\vp= r(i) \vp.
\end{equation}
This form is particularly helpful to write a mild solution to the Cauchy problem. We make the following assumption.
\begin{itemize}
\item [{\bf(A4)}] Let $\alpha:(0,\infty)^5\times\mc{X}\to [0,\infty)$ be such that $(s',v')\mapsto \alpha(s',v'; u,s,v,i)$ is the conditional probability density function of the random variable $(\hat{S}_u, \hat{V}_u)$ given $(\hat{S}_0, \hat{V}_0)=(s,v)$ and $X_{t'}=i$ for all $t'\in [0,u]$.\\
    (i) Then $\alpha\in C^{2,2,1,2,2}((0,\infty)^5\times\mc{X})$. \\
    Denote the first and second order partial derivatives of $\alpha$ w.r.t. $s$ by $\alpha_s$ and $\alpha_{ss}$ respectively. Similarly $\alpha_{u}, \alpha_{v},\alpha_{vv}, \alpha_{sv}$ denote other first and second order partial derivatives. Define $\mc{M}_{\alpha}(u,s,v,i):= \iint_{\mb{R}^2_+} s' \alpha(s',v'; u,s,v,i)ds'dv'$. Similarly define $\mc{M}_{\alpha_u}$, $\mc{M}_{\alpha_s}$, $\mc{M}_{\alpha_{ss}}$, $\mc{M}_{\alpha_v}$, $\mc{M}_{\alpha_{vv}}$ and $\mc{M}_{\alpha_{sv}}$.\\
    (ii) Assume that all seven maps defined above are finite valued.\\
    (iii) $\mc{M}_{\alpha_u}(\cdot,s,v,i)$ is continuous when other variables are kept fixed.\\
    (iv) The following maps  $\mc{M}_{\alpha}(\cdot,\cdot,\cdot,i)$, $\mc{M}_{\alpha_s}(u,\cdot,v,i)$, $\mc{M}_{\alpha_{ss}}(u,\cdot,v,i)$, $\mc{M}_{\alpha_v}(u,s,\cdot,i)$, $\mc{M}_{\alpha_{vv}}(u,s,\cdot,i)$ and $\mc{M}_{\alpha_{sv}}(u,\cdot,\cdot,i)$ are continuous uniformly in $u\in [0,T)$ when other variables are kept fixed.
\end{itemize}
\begin{rmk}
We note that for some special combinations of model parameters the above assumption can be verified very easily. For example, in the extreme case, if $r(i)=r$, a positive constant, and for all $i$, $\theta(i)=\hat{V}_0>0$ and $\sigma (i)=0$, then $\hat{V}$ is a constant process and $\hat{S}$ is a geometric Brownian motion. Therefore, $(s',v')\mapsto \alpha(s',v'; u,s,v,i)$ becomes a family of log-normal densities with mean and variance as continuous functions on $(u,s,v)$. Furthermore, all the partial derivatives of $\alpha$ can be expressed as products of $\alpha$ and sub-linear functions on $s'$. Thus (A4) (i-iv) hold.

Although (A4) appears desirable even for some non-trivial models, the verification of the same for an arbitrary combination of parameters may not be so easy. It seems, a comprehensive study on properties of $\alpha$ similar to (A4) is missing in the literature. However, $C^\infty$ regularity of $\alpha$ w.r.t. $s'$ variable appears in \cite{Rollin}.
\end{rmk}

\begin{thm}
\label{Cauchy}
Consider the Cauchy problem (\ref{main eqn.})-(\ref{2b}) and assume (A1), (A2)(i)-(iii) and (A4). Then
\begin{itemize}
  \item [(i)] the Cauchy problem has a unique classical solution $\vp$ with at most linear growth,
  \item [(ii)] $\vp$ is non-negative,
  \item [(iii)] the function $\frac{\partial\vp}{\partial s}$, partial derivative of solution w.r.t. $s$ variable, is bounded,
  \item [(iv)] $\sup_{i',y'}\sup_{\mc{D}}(\vp(t,s,v,i,y)- \vp(t,s,v,i',y'))$ is finite.
\end{itemize}
\end{thm}

\textbf{Proof:} {\bf (i)} Since $\hat{S}:=\{\hat{S}_t\}_{t\ge 0}$ has finite expectation (due to (A1)) and $K$ has at most linear growth, $K(\hat{S}_T)$ has finite expectation. Therefore, the mild solution $\vp$ to (\ref{3c}) and (\ref{2b}) can be written as
\begin{equation}\label{5}
\vp(t,s,v,i,y):= E[e^{-\int_t^Tr(X_u)du}K(\hat{S}_T)\mid \hat{S}_t=s, \hat{V}_t=v, X_t=i, Y_t =y].
\end{equation}
From (\ref{S.2}) we know at once that $\{ e^{-\int_0^t r(X_u)du} \hat{S}_t\}_{t\ge 0}$ is an $\{\mc{F}_t\}_{t\ge 0}$ martingale. Therefore,
$$E[e^{-\int_t^T r(X_u)du} \hat{S}_T | \hat{S}_t, \hat{V}_t, X_t, Y_t] = \hat{S}_t.$$
Now using the above relation, (\ref{5}), and the condition on $K$, we get
\begin{eqnarray}\label{6}
\no|\vp(t,s,v,i,y)-c_1s| &=&\bigg|E\left[e^{-\int_t^T r(X_u)du} K(\hat{S}_T)| \hat{S}_t=s,\hat{V}_t=v, X_t=i,Y_t=y\right]\\
\no&&- c_1 E\left[e^{-\int_t^T r(X_u)du} \hat{S}_T | \hat{S}_t=s,\hat{V}_t=v, X_t=x,Y_t=y\right]\bigg|\\
\no&\leq&  E\left[e^{-\int_t^T r(X_u)du}| K(\hat{S}_T)-c_1 \hat{S}_T |\mid \hat{S}_t=s,\hat{V}_t=v,X_t=i,Y_t=y\right]\\
&\leq&  c_2.
\end{eqnarray}
Thus $\vp$, as in  (\ref{5}) is a measurable function with at most linear growth.

Next the above function would be shown to satisfy an integral equation (IE). In order to derive the IE we need to introduce some notations. First we recall from Subsection 2.2, $n(t)= \max\{n\ge 0 \mid T_n\le t\}$. Thus $\{\omega \mid T_{n(t)+1}(\omega)>T\}$ represents the event of no transition during $[t,T]$ which is equal to $\{ \tau_{n(t)+1} > Y_t + (T-t)\}$ since $Y_t$ denotes the age of $X$ at the present state. By conditioning with $\tau_{n(t)+1}$, we write the mild solution (\ref{5}) as
\begin{equation}\label{ms}
 \vp(t,s,v,i,y) = E\left[E\left[e^{-\int_t^Tr(X_u)du}K(\hat{S}_T) \mid \hat{S}_t, \hat{V}_t, X_t, Y_t, \tau_{n(t)+1}\right] \mid \hat{S}_t=s, \hat{V}_t=v, X_t=i, Y_t =y\right].
\end{equation}
Consider $\tilde{S}, \tilde{V}$ satisfy (\ref{S.2}) only with exception that $X_t$ is replaced by $i$ at every place and the risk free asset price at $t$ is $e^{r(i)t}$. Define, $Hest(t,s,v,i):= E[e^{-r(i)(T-t)}K(\tilde{S}_T)\mid \tilde{S}_t=s, \tilde{V}_t=v]$. Therefore, the process  $\tilde{S}, \tilde{V}$ satisfy a classical Heston model and $Hest(t,s,v,i)$ is the price of $K(\tilde{S}_T)$ under a Heston model. Using the function $Hest$ we can rewrite (\ref{ms}) as follows
\begin{eqnarray*}
\lefteqn{\vp(t,s,v,i,y)}\\
&=& P[ \tau_{n(t)+1} > Y_t + (T-t)\mid \hat{S}_t=s, \hat{V}_t=v, X_t=i, Y_t =y] Hest(t,s,v,i)\\
 & &+ \int_0^{T-t} \frac{f(y+u\mid i)}{1-F(y \mid i)} E\left[e^{-\int_t^Tr(X_u)du}K(\hat{S}_T) \mid \hat{S}_t=s, \hat{V}_t=v, X_t=i, Y_t =y, \tau_{n(t)+1}=y+u\right]du.
\end{eqnarray*}
By further conditioning with random variables at time $t+u$, the above can again be rewritten as
\begin{eqnarray}\label{4}
\no \vp(t,s,v,i,y) &=& \frac{1-F(y+T-t\mid i)}{1-F(y \mid i)} Hest(t,s,v,i) + \int_0^{T-t} e^{-r(i)u} \frac{f(y+u\mid i)} {1-F(y \mid i)}\sum_{j \neq i}  p_{ij}(y+u)\times \\
&&  \int_{\mathbb{R}^2_+} \vp(t+u, s',v',j,0)\alpha(s',v'; u,s,v,i)ds'dv'du
\end{eqnarray}
where $\alpha$ is as in (A4). Thus we have proved that the mild solution (\ref{5}) satisfies the Volterra integral equation (\ref{4}). Next we would investigate (\ref{4}) for establishing sufficient regularity of (\ref{5}).

Let $U:= \{\vp: \mc{D}\to \mb{R} \textrm{ continuous} \mid \|\vp\|_U:= \sup_{\mc{D}} \frac{|\vp(t,s,v,i,y)|}{1+s} < \infty\}$ be the linear space of functions on $\mc{D}$ of at most linear growth in $s$ variable, with $\|\cdot \|_U$ as the norm. It is known that $(U, \|\cdot \|_U)$ is a Banach space. Note that, (\ref{4}) can be viewed as a fixed point problem of a contraction on $U$.
To prove this we follow the approach as in the Lemma 3.1 of \citep{AJP} which we partially derive for completeness. If we rewrite (\ref{4}) as $\vp =A\vp$, then for $\vp_1, \vp_2 \in U$
\be
\| A \vp_1 - A \vp_2 \| &\le& \sup_\mathcal{D} | \int_{0}^{T-t} e^{-r(i)u} \frac{f(y+u|i)}{1 - F(y|i)} \| \vp_1 - \vp_2 \| \frac{a(u,s,v,i)}{1+s} du |  \label{op}
\ee
where
\beo
a(u,s,v,i) &:=&  \int_{\mathbb{R}^2_+} (1 + s')\alpha(s',v';u,s,v,i) ds' dv' \n \\
&=& 1 + E\Bigr[\hat{S}_u \mid (\hat{S}_0, \hat{V}_0) = (s,v), X_t = i, 0 \le t \le u \Bigl].
\eeo
From (\ref{S.2}) we have
\beo
\hat{S}_u &=& \hat{S}_0 \exp\left( \int_{0}^{u} \Bigl( r(X_t)-\frac{1}{2}\hat{V}_t\Bigr) dt  + \int_{0}^{u} \sqrt{\hat{V}_t} d W_t^1 \right)= \hat{S}_0 \exp\left( \int_{0}^{u} r(X_t) dt\right) \mathcal{E} \left(\int_{0}^{u} \sqrt{\hat{V}_t} d W_t^1 \right)
\eeo
where the last multiplicative term is the Dol\'{e}ans-Dade exponential of $\int_{0}^{\cdot} \sqrt{\hat{V}_t} d W_t^1$. Thus the conditional expectation of $\hat{S}_u$ given that $X$ is constant in the interval $[0,u]$ and $(\hat{S}_0, \hat{V}_0) = (s,v)$ is equal to
\beo
 s e^{r(i)u} E\Bigr(\mathcal{E} \left(\int_{0}^{u} \sqrt{\hat{V}_t} d W_t^1 \right)| (\hat{S}_0, \hat{V}_0) = (s,v), X_t = i, 0 \le t \le u \Bigl)=  s e^{r(i)u}.
\eeo
Thus we can rewrite (\ref{op}) as
\beo
 \| A \vp_1 - A \vp_2 \| &\le& Q \| \vp_1 - \vp_2 \|,
\eeo
where
\be \label{QQ}
Q &:=& \sup_{\mathcal{D}} |\int_{0}^{T-t} e^{-r(i)u} \frac{f(y+u|i)}{1 - F(y|i)} \frac{1 + s e^{r(i)u}}{1+s} du |  <  \sup_{\mathcal{D}} |\int_{0}^{T-t} \frac{f(y+u|i)}{1 - F(y|i)} du |.
\ee
Now from Proposition \ref{prop} (iii)  we have that $F(y|i) = 1 - e^{- \int_{0}^{y} \Lambda_i(u)du}.$ From A2(i), $\Lambda_i(\cdot)$ is continuous on $[0,\infty)$. Hence it is bounded on every finite interval. Thus $e^{- \int_{0}^{y} \Lambda_i(u)du} >0$ for all $y >0$.  Or, in other words, $F(y|i)< 1$ for all $y >0.$ Therefore, $0 \le F(y+T-t|i) - F(y|i) < 1 - F(y|i)$. Hence
$$|\int_{0}^{T-t} \frac{f(y+u|i)}{1 - F(y|i)} du |= |\frac{F(y+T-t|i) - F(y|i)}{1 - F(y|i)}| < 1.
$$
Thus the right side of (\ref{QQ}) is less than or equal to 1. This proves that $Q$ is strictly less than 1, i.e., $A$ is a contraction operator on $U$.

Hence a direct application of Banach fixed point theorem assures existence of a unique continuous solution to (\ref{4}) with at most linear growth. Since (\ref{5}) is a measurable function with at most linear growth and satisfies (\ref{4}), the continuity of (\ref{5}) can be obtained if the image of (\ref{5}) under the contraction is continuous. In that case (\ref{5}) is the unique solution to (\ref{4}) in $U$.

There are two additive terms on the right of (\ref{4}). We discuss the improvement of regularity by considering only the second term on the right side of (\ref{4}), as the first term has desired regularity. Indeed from (A2)(i) and Proposition \ref{prop}(iii), $F(\cdot \mid i)$ is a $C^2$ function and the $Hest$ can be written as $Hest(t,s,v,i)=\int_{\mathbb{R}^2_+} K( s')\alpha(s',v'; T-t,s,v,i)ds'dv'$. Since $K$ has at most linear growth, the $C^{1,2,2}$ regularity of $Hest$, follows directly from assumptions (A4)(i-iv).

Note that the integrand of second term has bounded and continuous partial derivative w.r.t. $y$ variable which is integrated on a finite interval $[0,T-t]$. Thus the second term is continuously differentiable in $y$ variable. We invoke Assumption (A4) for establishing continuity and differentiability w.r.t. all other variables.

From (A2)(i), Proposition \ref{prop}(iii-v), and assumptions (A4)(i) we see that the integrand in the second term is a finite sum of products of $\vp$, the unknown, and a $C^1$ function in $u$ variable. Note that $\vp$, depends on $t$ variable as a function of $t+u$. Thus the partial differentiation of this unknown w.r.t. $t$ is as same as that w.r.t. $u$. Finally by using the integration by parts, at most linear growth of $\vp$, and the assumptions (A4) on $\mc{M}_{\alpha}$ and $\mc{M}_{\alpha_u}$, it is possible to transfer that partial derivative on the smooth coefficient and integrate w.r.t. $(s', v')$ to obtain a continuous function. Although the upper limit of the integral in second term also involves $t$ variable, but the integrand is continuous. Therefore $t$-dependency in the limit of integral causes no additional difficulty in proving regularity w.r.t. $t$ variable. Thus the second term is continuously differentiable in $t$.

The existence of continuous first and second order partial derivatives of the second term w.r.t. $s$ variable follow from at most linear growth of $\vp$ in $s'$ and (A4)(iv) assumption on $\mc{M}_{\alpha_s}$ and $\mc{M}_{\alpha_{ss}}$. Similarly, the assumptions in (A4)(iv) on $\mc{M}_{\alpha_v}$, $\mc{M}_{\alpha_{vv}}$ and $\mc{M}_{\alpha_{sv}}$ imply the existence of continuous partial derivatives $\frac{\partial}{\partial v}, \frac{\partial^2}{\partial v^2}$ and $\frac{\partial^2}{\partial s\partial v}$ of the second term.

Thus continuity of mild solution is established. Or in other terms, (\ref{5}) is indeed the unique solution to (\ref{4}) in $U$. Furthermore, (\ref{5}) also possesses $C^{1,2,2,1}$ regularity. Hence the mild solution is the classical solution. Thus (i) holds.

{\bf (ii)} The non-negativity follows from the non-negativity of $K$ in the expression of $\vp$ in (\ref{5}).

{\bf(iii)} Note that in the linear PDE (\ref{2a}) all the coefficients are smooth and the coefficients of the partial derivatives which are not with respect to $s$ are independent of $s$. In particular the coefficients of $\frac{\partial \vp}{\partial v}$ and $\frac{\partial^2 \vp}{\partial v^2}$ are independent of $s$. Thus the function $\frac{\partial \vp}{\partial s}$ also satisfies a linear parabolic PDE. Further, the terminal condition (\ref{2b}) is a Lipschitz continuous function in $s$ variable. This implies that the terminal condition for $\frac{\partial \vp}{\partial s}$ is bounded. Hence, the resulting Cauchy problem for $\frac{\partial \vp}{\partial s}$ will have a bounded solution. Or, in other words, $\frac{\partial \vp}{\partial s}$ is bounded. Therefore (iii) holds.

{\bf (iv)} From (\ref{6})  we get
\begin{eqnarray*}
\sup_{i',y'}\sup_{\mc{D}}|\vp(t,s,v,i,y)- \vp(t,s,v,i',y')| &=&\sup_{i',y'}\sup_{\mc{D}}|\vp(t,s,v,i,y)-c_1s +c_1s - \vp(t,s,v,i',y')|\\
   &\le& \sup_{i',y'}\sup_{\mc{D}}|\vp(t,s,v,i,y)-c_1s| + \sup_{i',y'}\sup_{\mc{D}}|\vp(t,s,v,i',y')- c_1s|\\
   &\le& 2c_2
\end{eqnarray*} which is finite.\qed
\begin{rmk}
The solution to equation (\ref{main eqn.})-(\ref{2b}) does not have a closed form expression. Hence one should find a numerical solution to this. A direct approach involves finite difference method. It is not difficult to develop a Crank-Nicolson type stable implicit scheme for the same. However, there are some other indirect approaches as well which we discuss below.

In the above proof we have showed that the mild solution (\ref{5}) is the unique continuous solution to (\ref{4}) and that has desired smoothness to become the classical solution to (\ref{main eqn.})-(\ref{2b}). Hence (\ref{main eqn.})-(\ref{2b}) and (\ref{4}) are equivalent. Hence one can obtain a numerical solution to (\ref{4}) by employing a quadrature method to solve numerically.

There is another indirect numerical approach which involves direct computation of (\ref{5}). Let $\bar{\alpha}:(0,\infty)^7 \times\mc{X}\to [0,\infty)$ be such that $(s',v')\mapsto \bar{\alpha}(s',v',u,s,v,i,y,t)$ be the conditional probability density function of the random variable $(\hat{S}_u,\hat{V}_u)$ as defined in (\ref{S.2}) given $(\hat{S}_t = s, \hat{V}_t = v, X_t = i, Y_t = y)$ for some $0 \le t < u$. Then following the methods as in the proof of Theorem \ref{Cauchy} it can be shown that $\bar{\alpha}$ satisfies the integral equation:
\beo
\lefteqn{\bar{\alpha}(s',v',u,s,v,i,y,t)}\\
&=& \frac{1 - F(y+u -t \mid i)}{1 - F(y \mid i)} \alpha(s',v',u-t,s,v,i)  \\
&& + \int_{0}^{u-t} \int_{\mb{R}_2^+} \sum \limits_{j \neq i} \bar{\alpha}(s',v',u,\bar{s}, \bar{v}, j,0, t + \bar{u}) \alpha(\bar{s}, \bar{v}, \bar{u}, s,v,i) p_{ij}(y + \bar{u}) \frac{f(y + \bar{u} \mid i)}{1 - F(y \mid i)} d \bar{s} d\bar{v} d\bar{u}
\eeo
where $\alpha$ is as in (A4). The above equation can be used to compute the conditional density function $\bar{\alpha}$ when $\alpha$ is known. If $r$ is a constant, then $\bar{\alpha}$ can directly be used to find the conditional expectation (\ref{5}). This gives rise to another numerical method to compute the solution to the Cauchy problem (\ref{main eqn.})-(\ref{2b}).
\end{rmk}

\section{Derivation of F\"ollmer Schweizer decomposition}
\label{derivation}

In this section we find the F\"ollmer-Schweizer decomposition of the discounted terminal payoff $\frac{1}{B_T} K(S_T)$ where $S$ and $B$ are as in Subsection 2.1. Or in other words, we would show that there is an $\{\mc{F}_t\}$-adapted process $\xi=\{\xi_t\}_{t\ge 0}$ such that
$$ \frac{1}{B_T} K(S_T) = H_0 + \int_{0}^{T} \xi_t d S_t^* + L_T,$$
where $H_0$ is $\mc{F}_0$-measurable and  $L:= \{ L_t:= E(L_T|\mc{F}_t) \}$ is a mean-zero square integrable martingale and orthogonal to $M:= \{M_t\}_{t\ge 0}$, the martingale part of $S$, given by
\begin{equation}\label{M}
M_t:=\int_0^t S_u\sqrt{V_u} dW^1_u.
\end{equation}
It is known (see \cite{S2}) that the integrand $\xi$ constitutes the optimal hedging of the contingent claim $K(S_T)$ and the locally risk minimizing price at time zero of the claim is given by $H_0$. Therefore, the problem of pricing and hedging boils down to finding an appropriate choice of $\xi$ and $H_0$. Generally, precise expressions of these quantities for European type path independent options involve solution and its partial derivatives of a relevant Black-Scholes-Merton type equation. Of course, the exact form of the equation depends on the asset price dynamics.

In this section we prove that the classical solution obtained in preceding section is indeed the local risk minimizing option price. We make the following assumption.
\begin{itemize}
  \item [{\bf(A5)}] If $\vp$ denotes the unique classical solution to (\ref{main eqn.})-(\ref{2b}), then
$$E\int_0^T V_t\left(\frac{\partial\vp}{\partial v} (t,S_t, V_t, X_{t-}, Y_{t-})\right)^2 dt <\infty.$$
\end{itemize}
\begin{rmk} Note that the above condition asserts the square integrability of the Greek Vega of the option with claim $K(S_T)$. Under classical B-S model the Vega of call and put options are square integrable.
\end{rmk}

\begin{thm}\label{th1}
Let $\vp$ be the unique classical solution to the Cauchy problem (\ref{main eqn.})-(\ref{2b}) with at most linear growth. Let $(\xi, \varepsilon)$ be given by $\xi_t:= \Big(\frac{\partial\vp}{\partial s}+ \frac{\rho \sigma(X_t)}{S_t} \frac{\partial\vp}{\partial v}\Big)(t,S_t, V_t, X_{t-}, Y_{t-})$ and $\varepsilon_t := e^{-\int_0^tr(X_u)du} (\vp(t,S_t, V_t, X_{t-}, Y_{t-})-\xi_tS_t)$. Then under (A1), (A2), (A3), (A4) and (A5)
\begin{itemize}
  \item[(i)]  $E\int_{0}^{T} \xi^2_t d \langle S\rangle_t < \infty$,
  \item[(ii)] $E(\varepsilon_t^2) < \infty$ for all $t\ge 0$,
  \item[(iii)] $L_T:=  \frac{1}{B_T} K(S_T) -\vp(0,S_0, V_0, X_0, Y_0)- \int_{0}^{T} \xi_t d S_t^*$ belongs to $L^2(P)$,
  \item[(iv)] $L:= \{L_t=E(L_T|\mc{F}_t) \} $ is orthogonal to $M$ as in (\ref{M}).
  \item[(v)]  $(\xi, \varepsilon)$ is the optimal hedging strategy,
  \item[(vi)]  $\vp(0,s,v, i,y)$ is the locally risk minimizing price at time zero of the European option with terminal payoff $K(S_T)$, when $S_0=s, V_0=v, X_0=i, Y_0 =y$.
\end{itemize}
\end{thm}


\textbf{Proof:}
{\bf (i)} For the ease of notation, we write $p_t$ to denote $(t,S_t, V_t, X_{t-}, Y_{t-})$ and $p_0= (0,S_0, V_0, X_{0}, Y_{0})$. Note that
\begin{eqnarray*}
  \int_{0}^{T} \xi^2_t d \langle S\rangle_t &=& \int_{0}^{T} \Big(\frac{\partial\vp}{\partial s}+ \frac{\rho \sigma(X_t)}{S_t} \frac{\partial\vp}{\partial v}\Big)^2(p_t) d \langle S\rangle_t  \\
   &\le & 2\int_{0}^{T} \Big(\frac{\partial\vp}{\partial s}(p_t)\Big)^2 S_t^2V_t dt   +  2 \rho^2 \int_{0}^{T} \sigma^2(X_t) V_t  \Big(\frac{\partial\vp}{\partial v}(p_t)\Big)^2 dt.
\end{eqnarray*}
Using square integrability of $S$ (this follows from (A1)) and boundedness of $\frac{\partial\vp}{\partial s}$ ( see Theorem \ref{Cauchy}(iii)), we conclude that the first integral has finite expectation. Again finiteness of expectation of the second integral is due to the assumption (A5).

{\bf (ii)} We note that $|\varepsilon_t| \le |\vp(t,S_t, V_t, X_{t-}, Y_{t-})-\xi_t S_t|$. Using at most linear growth of $\vp$ (see Theorem \ref{Cauchy}(i)), and square integrability of $S$, expectation of square of $\vp(t,S_t, V_t, X_{t-}, Y_{t-})$ is finite. Again (A3) ensures existence of moment generating function of $\int_{0}^{T} \frac{1}{V_s} ds$ in a neighborhood of zero. Hence $E\int_{0}^{T} \frac{1}{V_s} ds$ is finite. Therefore, this and (i) together imply that $E[(\xi_t S_t)^2]$ is also finite. Thus (ii) holds.

{\bf (iii)} This directly follows from (i), the fact that $B_T$ is bounded away from zero and $K$ has at most linear growth.

{\bf (iv)} We rewrite equation (\ref{2a}) using the infinitesimal generator $\mc{L}$ of the Markov process $\{(S_t, V_t, X_t, Y_t)\}_{t\ge 0}$ in the following way
\be\label{2c}
\left( \frac{\partial}{\partial t} + \mc{L} - r(i)\right)\vp  = (\mu(i) - r(i))s \frac{\partial\vp}{\partial s}  +  \rho \sigma(i)(\mu(i) - r(i)) \frac{\partial\vp}{\partial v}.
\ee
If $\vp$ solves (\ref{2a})-(\ref{2b}) classically, using It\={o}'s formula, we have from equation (\ref{Heston})-(\ref{Y})
\be \label{3}
\no \lefteqn{d\Bl \frac{1}{B_t} \vp(t,S_t, V_t, X_t, Y_t)\Br} \\
\no &=& -r(X_{t-})\frac{1}{B_t} \vp(p_t)dt + \frac{1}{B_t} \Bl \frac{\partial}{\partial t} + \mc{L} \Br \vp(p_t) dt  + \frac{1}{B_t} \Big[ S_t \sqrt{V_t} \frac{\partial\vp}{\partial s}(p_t) d W^1_t + \sigma(X_t) \sqrt{V_t} \frac{\partial\vp}{\partial v}(p_t)  dW_t^2 \\
&& + \int_{\mb{I}} \Bl \vp(p_t + (0,0,0, h_{\lambda}(X_{t-}, Y_{t-},z),- g_{\lambda}(X_{t-},Y_{t-},z))) - \vp(p_t) \Br \tilde{\wp}(dt,dz) \Big],
\ee
where $\tilde{\wp}(dt,dz)$ is the compensated Poisson random measure given by $\tilde{\wp}(dt,dz) := \wp(dt,dz) - dtdz.$ Integration on both sides of (\ref{3}) from 0 to $T$ w.r.t. $t$ and use of the terminal condition yield
\be
\frac{K(S_T)}{B_T} &=& \vp(p_0) + \int_{0}^{T} \frac{1}{B_t}\Big( \frac{\partial}{\partial t}  + \mc{L} - r(X_t) \Big) \vp(p_t) dt + \int_{0}^{T} S^*_t \sqrt{V_t} \frac{\partial\vp}{\partial s}(p_t) d W^1_t  + \int_{0}^{T} \frac{\sigma(X_t) \sqrt{V_t}}{B_t} \frac{\partial\vp}{\partial v}(p_t) d W^2_t \n \\
&& + \int_{0}^{T} \frac{1}{B_t} \int_{\mb{I}} \Bl \vp(p_t + (0,0,0, h_{\lambda}(X_{t-}, Y_{t-},z),- g_{\lambda}(X_{t-},Y_{t-},z))) - \vp(p_t) \Br \tilde{\wp}(dt,dz) \label{1}
\ee
where  $S_t^* := \frac{S_t}{B_t}$. Since $\vp$ solves (\ref{2c}), the first integral on the right of (\ref{1}) can be rewritten to obtain
\beo
\frac{K(S_T)}{B_T} &=&  \vp(p_0) + \int_0^T \Bigl( \mu(X_t) - r(X_t) \Bigr)S_t^* \frac{\partial\vp}{\partial s}(p_t) dt + \rho \int_{0}^{T} \frac{\sigma(X_t)}{B_t} \Bigl( \mu(X_t) - r(X_t) \Bigr)  \frac{\partial\vp}{\partial v}(p_t) dt  \n \\
&& +  \int_{0}^{T} S_t^* \sqrt{V_t}  \frac{\partial \vp}{\partial s}(p_t)dW_t^1 + \int_{0}^{T} \frac{{\sigma(X_t) \sqrt{V_t}}}{B_t} \frac{\partial\vp}{\partial v}(p_t) dW_t^2 \n \\
&&+ \int_{0}^{T} \frac{1}{B_t}  \int_{\mb{I}} \Bl \vp(p_t + (0,0,0, h_{\lambda}(X_{t-}, Y_{t-},z),- g_{\lambda}(X_{t-},Y_{t-},z))) - \vp(p_t) \Br \tilde{\wp}(dt,dz).
\eeo
By using $dS_t^* = S_t^* \Bl ( \mu(X_t) - r(X_t) )dt + \sqrt{V_t} dW_t^1 \Br $ and combining the second and the fourth additive terms on the right side of the above equation, we have
\be
\frac{K(S_T)}{B_T} &=& \vp(p_0) + \int_{0}^{T} \frac{\partial\vp}{\partial s}(p_t) d S_t^* + \rho  \int_{0}^{T} \sigma(X_t)\frac{1}{S_t} \frac{\partial\vp}{\partial v}(p_t) (dS_t^*- S_t^* \sqrt{V_t} dW_t^1) + \int_{0}^{T} \frac{\sigma(X_t) \sqrt{V_t}}{B_t} \frac{\partial\vp}{\partial v}(p_t) dW_t^2 \n \\
&& + \int_{0}^{T} \frac{1}{B_t} \int_{\mb{I}} \Bl \vp(p_t + (0,0,0, h_{\lambda}(X_{t-}, Y_{t-},z),- g_{\lambda}(X_{t-},Y_{t-},z))) - \vp(p_t) \Br \tilde{\wp}(dt,dz).
\label{eq.2}
\ee
Now we rearrange the terms on right hand side of (\ref{eq.2}) to get
\be
\frac{K(S_T)}{B_T} &=& \vp(p_0) +  \int_{0}^{T} \xi_t dS_t^* + \int_{0}^{T} \frac{\sigma(X_t) \sqrt{V_t}}{B_t}  \frac{\partial\vp}{\partial v}(p_t) \Bl d W^2_t - \rho dW_t^1\Br \n \\
&+& \int_{0}^{T} \frac{1}{B_t} \int_{\mb{I}} \Bl \vp(p_t + (0,0,0, h_{\lambda}(X_{t-}, Y_{t-},z),- g_{\lambda}(X_{t-},Y_{t-},z))) - \vp(p_t) \Br \tilde{\wp}(dt,dz).
\label{FS decom}
\ee
Hence $L$ as in (iii)-(iv) is equal to
$$\int_{0}^{\cdot} \frac{\sigma(X_t) \sqrt{V_t}}{B_t}  \frac{\partial\vp}{\partial v}(p_t) \Bl d W^2_t - \rho dW_t^1\Br + \int_{0}^{\cdot} \frac{1}{B_t} \int_{\mb{I}} \Bl \vp(p_t + (0,0,0, h_{\lambda}(X_{t-}, Y_{t-},z),- g_{\lambda}(X_{t-},Y_{t-},z))) - \vp(p_t) \Br \tilde{\wp}(dt,dz).
$$

Since for any $t>0, \ W_t^2 - \rho W_t^1$ is orthogonal to $W_t^1$, the first term using (A5) is a $L^2$ martingale which is orthogonal to $M$. As $\tilde{\wp}$ is independent of $W^1_t$, using Theorem \ref{Cauchy} (iv) with finiteness of $\mb I$, the second term, being an integral of a bounded predictable process w.r.t. compensated random measure $\tilde{\wp}$ is also a square integrable martingales orthogonal to $M$. Hence (iv) follows.

{\bf (v)} To justify (v), we show that $(\xi, \varepsilon)$ fulfills all three conditions given in Definition \ref{def1}. The first two square integrability conditions are already shown above. Using the definition of $\varepsilon$, the remaining condition translates to checking if there is a constant $a$ such that $P(\vp(p_t)\geq -a,~\forall~ t)=1$. Using Theorem \ref{Cauchy}(ii) asserts that $\vp$ is nonnegative. Hence the above condition holds with $a=0$.

{\bf(vi)} This follows from the F\"{o}lmer-Schweizer decomposition (\ref{FS decom}) obtained above. \qed

\section{Conclusion}
In this paper we deal with the problem of option pricing for a European call option when the underlying stock price follow a semi-Markov modulated Heston model. This model improves upon the existing semi-Markov modulated models in the literature, for example \citep{AJP}, where the stock volatility is a finite-state pure jump process.

It should be noted that the derivations in this paper is different from the standard approach, for example, \citep{SWH}. Here we start with a Cauchy problem which we show to possess a classical solution. We then construct a hedging strategy using the first order partial derivatives of the solution so as to obtain F\"ollmer Schweizer decomposition of contingent claim related to a European option. From the decomposition we conclude that the solution to the Cauchy problem is indeed the locally risk minimizing price of the corresponding European option. This approach avoids an a-priori tacit assumption of desired differentiability of the option price function that is expressed using a conditional expectation with respect to an equivalent minimal martingale measure. We have also obtained an integral equation of option price.

To the best of our knowledge this is the first work on a model where the parameters of a stochastic volatility model change regime according to a semi-Markov process. We hope that this work will open up newer extensions in the future.

{\bf Acknowledgement: } We acknowledge Anup Biswas for some useful discussion. We are also thankful to the anonymous referee and the associate editor for some very useful suggestions.

\end{document}